\documentstyle[11pt]{article}
\begin{document}
\begin{flushright}
{\bf{\large TPJU 12/95}}
\end{flushright}
\begin{flushright}
  {\bf{\large hep-ph-9506236}}
\end{flushright}
\begin{flushright}
  {\bf June 1995}
\end{flushright}
\vskip 20pt
\begin{center}
{\huge\bf The Role of Chromo-magnetic Vacuum Background Field
in $e^{+}e^{-}\rightarrow 2\  jets$ and other reactions\footnote{sponsored
 in part by the KBN Grant No 2-P03B-083-08}}
\end{center}
\begin{center}
{\large W.Czy\.{z}\footnote{e-mail address czyz@ztc386a.if.uj.edu.pl}}
\end{center}
\begin{center}
  {\small Institute of Physics, Jagellonian University }
\end{center}

\begin{center}
  {\small Reymonta 4, 30-059 Krak\'{o}w, Poland}
\end{center}

\begin{center}
  {\small and}
\end{center}
\begin{center}
  {\small Institute of Nuclear Physics}
\end{center}

\begin{center}
  {\small Radzikowskiego 152, 31-342 Krak\'{o}w, Poland}
\end{center}

\begin{center}
  {\large and}
\end{center}

\begin{center}
  {\large J.Turnau\footnote{e-mail address turnau@chall.ifj.edu.pl}}
\end{center}

\begin{center}
  {\small Institute of Nuclear Physics}
\end{center}

\begin{center}
  {\small Kawiory 26a, 30-055 Krak\'{o}w, Poland}
\end{center}
\vskip 10pt
\begin{abstract}
We propose a new type of a measurement which is sensitive to the QCD
vacuum color-magnetic  fluctuations: A measure  of the axial asymmetry
of the hadronic final states produced in the high energy  $e^{+}e^{-}$
collisions which is related to the chromomagnetic vacuum field strength.
\end{abstract}
\vskip 10pt
\section{Introduction}
\par
In this note we propose a new type of a measurement, sensitive
to the presence of the chromomagnetic vacuum background field \---
an idea with almost 20 years of history [1] (see [2] for a recent
review).
\par
The basic idea of the effect we discuss here has been proposed more than 10
years ago by O.Nachtmann and A.Reiter [3] : the chromomagnetic vacuum field
changes the  trajectory of partons created in the high energy collisions,
giving rise to various correlations between the parton and, indirectly,
the hadron
momenta. Although the possibility to investigate such a basic property of QCD
 seems
to be very attractive, no major effort
has been done in this direction for the two important reasons:
\begin{itemize}
\item
these effects, if exist, are very difficult to measure, and
\item
it is not obvious that they could not be
interpreted in terms of the perturbative QCD theory, i.e. there might be
serious problems with a unique interpretation of such effects.
\end{itemize}
A good example of the second point is the K-factor in the Drell-Yan process,
 which
can be interpreted either as the result of the color polarization of
 quarks traversing
the domain of the vacuum chromo-magnetic field [3] or, simply, as the higher
 order QCD corrections.
\par
In recent publications by Nachtmann {\it et al.} [4] some arguments
in favor of observability of the chromomagnetic field of the vacuum  have
been reiterated, refined, and some supporting experimental evidence quoted.
On the theoretical  side, some support for the concept of
a ferromagnetic vacuum comes also from the recent
lattice simulations [5].
\par
A very large statistics of $e^{+}e^{-}$ annihilations into hadrons,
collected in LEP and SLD experiments encourages us to propose a new type
of a measurement  designed to detect deflection of
 quarks  and antiquarks in the chromomagnetic fields of the QCD vacuum.
Its new feature is a possibility to distinguish the true QCD vacuum effect from
 the perturbative interpretations based on the
CP-parity conservation arguments.
Observation of CP-parity violating  effects would provide a strong support
for the non-perturbative vacuum structure of QCD which,
 indeed, may locally violate CP-parity.
\par
Our discussion is mainly devoted to the $e^{+}e^{-}$ annihilations
 (sections 2 and 3) where, from
the phenomenological point of view, situation is the simplest. However the
hadron-hadron and, above all, nucleus-nucleus collision data, may also provide
very interesting evidence for the structured non-perturbative QCD vacuum.
 Such effects are
briefly discussed in section 4. Section 5 contains the summary and the
conclusions.
\section {Quarks in the vacuum background chromomagnetic field}
Following references [3] and [4],
 let us consider QCD vacuum as a sort of  ``ether''. At any given time it is
supposed to show a domain structure. Inside the domains there are more or less
constant chromomagnetic fields. Theory does not have much to say about the
space
extension and the time duration of such vacuum fluctuations. Seemingly a
natural assumption [3] is the linear dimension of the domain of order
$1/\Lambda$ and the frequency of fluctuations of order $\Lambda$, where
$\Lambda$ is the QCD scale parameter. However we are aware that simple
intuition might be not the best guide in this case.
 The reasoning which follows has been inspired by the model of the  origin
 of the jet handedness by Ryskin [6].
\par
At the stage of  fragmentation of a color-electric
string, spanned between the initial $(q\bar{q})_{i}$ pair created
in $e^{+}e^{-}$ collision, the secondary $(q\bar{q})_{s}$ pairs are
created. Let us discuss a very simplified picture of
this breakup: all quarks have the same color and antiquarks have the same
anti-color. Thus, in a domain of the constant chromomagnetic vacuum field,
 all quarks turn, in the plane transverse to the string,
in one direction and antiquarks in the opposite direction.
\par
Let us consider trajectories of secondary $(q\bar{q})_{s}$ pairs in
more detail.
At the beginning, the transverse (with
respect to the common  axis of jets $\vec{l}$) momenta of these new partons
are balanced
\begin{displaymath}
\vec{q}_{\perp}= - \vec{\bar{q}}_{\perp}
\end{displaymath}
but, when moving in the the color magnetic field, they acquire equal,
 additional
momenta $\delta\vec{q}_{\perp}=\delta\vec{\bar{q}}_{\perp}$
(as they have opposite color charges and move in opposite directions in the
transverse plane). As the result we get a triad of vectors
\begin{equation}
\vec{q}_{\perp}^{\ '}=\vec{q}_{\perp} + \delta\vec{q}_{\perp };
\vec{\bar{q}}_{\perp}^{\ '}=\vec{\bar{q}}_{\perp} + \delta\vec{q}_{\perp };
\vec{l}
\end{equation}
with the handedness i.e. the sign of the mixed product
$(\vec{q}_{\perp }^{\ '} \times \vec{\bar{q}}_{\perp }^{\ '})\cdot \vec{l}$
uniquely defined by
 the relative orientation of $\vec{l}$ and the background chromomagnetic
 field $\vec{B}$.
{\bf In a constant vacuum field all triads (in both jets) have the same
handedness}.
\par
For comparison, in the Ryskin model [6] the   chromomagnetic fields are
 produced
 by the color-dipole moments of the original quark and anti-quark. The
direction
of these fields and thus the
 relative handedness  of triads (1)
 in the opposite jets depends on the spin alignment for the original
$(q\bar{q})_{i}$ pair, but is always CP-symmetric.
In the particular case of the
aligned spins, as in the case of $e^{+}e^{-}$ collisions, {\bf
handedness produced
by color-dipole chromomagnetic fields is opposite for the quark and anti-quark
jets}. As will be discussed later the effect of a constant vacuum
 chromomagnetic  field can
be easily distinguished  from the other QCD effects merely by observing
 the {\bf "wrong sign" handedness correlation} between jets.
\par
 The value of  the additional
transverse momentum $\mid\delta\vec{q}_{\perp}\mid$ acquired by the parton
 can be roughly related to the strength of the vacuum chromo-magnetic field
assuming that the parton before loosing its color charge  traverses
a typical hadronization distance $1/\Lambda$ [3]. Then
\begin{equation}
\mid\delta \vec{q}_{\perp}\mid=\sqrt{\alpha_{s}}\langle
\mid(\vec{B}\cdot\vec{l})\mid\rangle/\Lambda
\end{equation}
It is interesting to note that $\mid\delta\vec{q}_{\perp}\mid$  may be quite
substantial. One parameter characterizing the vacuum fields is the expectation
value of the square of the gluon field strength introduced by Vainstein
{\it et al.} [7].
 Using the experimental value given in the  recent review of non-perturbative
methods in QCD [8],
\begin{displaymath}
  \langle0\mid 4\pi\alpha_{s}\vec{B}^{a}\vec{B}^{a}\mid 0 \rangle\approx
(700 MeV)^{4}
\end{displaymath}
we have
$\sqrt{\alpha_{s}}\langle \mid(\vec{B}\cdot\vec{l})\mid\rangle /\Lambda
 \approx 200MeV$
 i.e. a quantity comparable
to the average parton transverse momentum $(\approx 400 MeV)$!
Thus, at the parton level the axial asymmetry induced by the vacuum field
may be large enough but, of course, it is washed out to a large extent (if not
completely) after the hadronization.
\section{Transfer of the  axial asymmetry from parton to hadron level}
\par
Transfer of the axial asymmetry from the parton level to the hadron level is
possible, at least in principle, due to the local retention of the
parton quantum numbers in the hadronization process. A possibility of
 discovering of such
an asymmetry may crucially depend on the proper choice of the kinematical
variables which we employ in our analysis of the hadronic final state.
\par
The "jet handedness" discussed in the literature [9,10,11] is an example
 of such
quantity. Its notion was originally connected to the idea that axial
asymmetry due to the
polarization of the quark from which jet originates, may be induced into the
jet as a quantum interference effect [10] or color-dipole field effect [6].
Apparently, such effects have to be {\bf local in the rapidity space} and the
quantities designed to measure it have to be based on local variables.
Typically one chooses [11] the oppositely charged tracks of the highest
 momentum and
from their momenta $\vec{k}_{+}$, $\vec{k}_{-}$ one constructs quantity
(related  to the jet handedness)
\begin{equation}
  \omega=\vec{t}\cdot(\vec{k}_{+} \times  \vec{k}_{-}),
\end{equation}
where $\vec{t}$ is the thrust axis in  the jet direction.
A signal would be visible as a nonzero mean $\langle\omega\rangle$,
 proportional to the polarization of the initial quark.
\par
Incidentally, in the recent data from SLD [11] $\langle\omega\rangle$
 was found to be consistent with zero.
 The correlation  between
$\omega_{1}$ and $\omega_{2}$ in the opposite jets:
    \begin{equation}
R_{\omega}=(\langle\omega_{1}\omega_{2}\rangle-\langle\omega_{1}\rangle
\langle\omega_{2}\rangle )
    \end{equation}
was for the first time considered by Efremov, Potashnikova and Tkatchev [12],
who observed a correlation signal in DELPHI preliminary data. In SLD data [11]
no jet handedness correlation was observed, however a large difference in
the statistics of these two experiments should be noted.
\par
In contrast, in our case
the {\bf asymmetry induced into parton state by the interaction with the vacuum
background chromomagnetic field has global character} i.e. it is distributed
over the whole interaction volume. So , we propose to build a measure of
the jet handedness from cumulative variables. At first, we define cumulant of
 the transverse momentum for positive and negative particles separately
 \begin{equation}
 \vec{P}_{\perp}^{\pm}(y_{min},y_{max})= \sum_{j}\vec{k}_{\perp j}
\Theta(y_{max}-y_{j})\Theta(y_{j}-y_{min})\Theta(\pm Q_{j})
 \end{equation}
where $\vec{k}_{\perp j},y_{j},Q_{j}$ denote the  transverse momentum,rapidity
and charge of the j-th particle,
$\Theta(x)$ is the step function equal to 1(0) for $x>0(x<0)$. This quantity
is the sum of the transverse momenta of all positive (negative) particles
in the rapidity range $(y_{min},y_{max})$.
\begin{equation}
P_{\perp }(y_{min},y_{max})=\mid (\vec{P_{\perp }}^{+}(y_{min},y_{max})+
\vec{P_{\perp }}^{-}(y_{min},y_{max}))\mid
\end{equation}
is the length of the total transverse momentum vector in the above rapidity
range.
\par
The standard assumption of the local compensation of transverse momentum,
equivalent to the assumption that $(q\bar{q})_{s}$ pairs do not carry
momentum transverse to the jet axis,
leads to the prediction, that far from the phase space boundary
$P_{\perp}$ remains independent of $\Delta y=\mid y_{min}-y_{max}\mid$
i.e. does not depend on the number
of $q\bar{q}$ pairs created in the rapidity range $(y_{min},y_{max})$.
 However, in the model with the background
chromomagnetic field the situation is different:
each  $(q\bar{q})_{s}$ pair acquires the
transverse momentum $2\mid\delta\vec{ q}_{\perp }\mid$.
 It is randomly oriented in the plane
transverse to the jet direction, thus we expect that
the average value of the transverse momentum cumulant grows proportionally to
 the square root of the average number $N_{q\bar{q}}$ of
$(q\bar{q})_{s}$ pairs in the rapidity range $(y_{min},y_{max})$:
\begin{equation}
\langle P_{\perp }(y_{min},y_{max})\rangle \approx \sqrt{ N_{q\bar{q}}}
\mid\delta \vec{q}_{\perp }\mid=
 \sqrt{\overline{n} \mid y_{min}-y_{max}\mid} \mid\delta\vec{ q}_{\perp}\mid,
\end{equation}
where $\bar{n}$ is the average number of $q\bar{q}$ pairs per unit
rapidity interval and $\langle\ \rangle$ denotes an average over events.
In other words the transverse momentum shows the
effect of diffusion.
\par
Employing the cumulative variables $\vec{P}_{\perp}^{\pm}$ we can construct
 the mixed
product analogical to that in eq (3):
\begin{equation}
  \Omega(y_{min},y_{max})=(\vec{P}_{\perp}^{+}(y_{min},y_{max}) \times
\vec{P}_{\perp}^{-}(y_{min},y_{max}))\cdot \vec{l},
\end{equation}
where $\vec{l}$ is the direction along the common axis of the two jets. Its
sign is related to the handedness of triads (1). Note that we do not use vector
$\vec{t}$ oriented in the jet direction, as in formula (3), in order to avoid
a singularity at the rapidity $y=0$, so that we can integrate
 over the rapidity  range which includes this point.
We can define also the correlation function $R_{\Omega}$
    \begin{equation}
R_{\Omega}=(\langle\Omega_{1}\Omega_{2}\rangle-\langle\Omega_{1}\rangle
\langle\Omega_{2}\rangle ),
    \end{equation}
where $\Omega_{1}=\Omega(y_{min},y_{max})$ and
 $\Omega_{2}=\Omega(-y_{min},-y_{max})$.
This quantity is, in turn, analogical to $R_{\omega}$ defined in eq (4).
\par
As we have seen before, the handedness of triads (1)
 induced by the chromomagnetic vacuum background field in the opposite jets
is the same .
Let us see how it is transferred to the quantity $\Omega$ defined above.

The assumptions of the local retention of the parton quantum numbers,
of the momentum conservation and the isospin symmetry lead to the  approximate
expressions in terms of the quark momenta
\begin{equation}
  \vec{P}_{\perp}^{+}(y_{min},y_{max})\approx
\frac{1}{3}(1+\beta)\sum_{i}\vec{q}_{\perp i}+
                   \frac{1}{3}(1-\beta)\sum_{i}\vec{\bar{q}}_{\perp i},
\end{equation}
\begin{equation}
  \vec{P}_{\perp}^{-}(y_{min},y_{max})\approx
\frac{1}{3}(1-\beta)\sum_{i}\vec{q}_{\perp i}+
                   \frac{1}{3}(1+\beta)\sum_{i}\vec{\bar{q}}_{\perp i}.
\end{equation}
$\sum_{i}$ extends over all $(q\bar{q})_{s}$ pairs falling into the
 rapidity
range $(y_{min},y_{max})$. It should be noted that relations (10) and (11)
are applicable for both jets only in  the central rapidity range, where
parton states in both jets are neutral in charge and flavor.The situation is
much less clear in the quark (anti-quark) fragmentation regions \---
we will come back to this problem later.
\par
 The parameter $\beta$ in the above formulas can be related to the
experimentally  measured  quantity
\begin{equation}
  \langle\Delta P_{\perp}\rangle=
\langle\mid \vec{P}_{\perp}^{+}-\vec{P}_{\perp}^{-}\mid
\rangle\approx \frac{4}{3}\beta\langle\mid\sum_{i}\vec{q}_{\perp i}\mid\rangle
\approx  \frac{4}{3}\beta\sqrt{N_{q\bar{q}}}\langle
q_{\perp}\rangle
\end{equation}
where, in analogy with formula (7), we have used the argument about the
 random walk in the transverse momentum plane to get
\begin{equation}
\langle\mid\sum_{i}\vec{q}_{\perp i}\mid\rangle \approx
\sqrt{N_{q\bar{q}}}\langle q_{\perp}\rangle.
\end{equation}
Note that this formula does not contain $\vec{B}$ because of cancellation of
the $\delta\vec{q}_{\perp i}$ components in (12).
 From (2), (8), (10), (11) and (12)
we get
\begin{equation}
  \Omega(y_{min},y_{max})\approx \Omega(-y_{min},-y_{max})
\approx\frac{8}{9}\beta \sqrt{N_{q\bar{q}}}
\sqrt{\alpha_{s}}\langle q_{\perp}\rangle(\vec{B}\cdot\vec{l})/\Lambda.
\end{equation}
The average  $\langle\Omega\rangle$ vanishes, as for the randomly oriented
chromomagnetic vacuum field  $\langle(\vec{B}\cdot\vec{l})\rangle=0$.
The  correlation function $ R_{\Omega}$ has a
non zero value
\begin{equation}
  R_{\Omega}(y_{min},y_{max})\approx
\frac{4}{9}\langle\Delta P_{\perp}\rangle^2
\alpha_{s}\langle \mid(\vec{B}\cdot\vec{l})\mid\rangle^{2}/\Lambda^{2}.
\end{equation}
where we employed the formula  (12) to  eliminate $\beta N_{q\bar{q}}$, so that
$ R_{\Omega}$ is  related to the vacuum field strength only
through the parameters which can be measured in the same experiment.
\par
The sign of the correlation function is positive and in this respect our effect
differs from Ryskin  approach [6] where, by the requirement of
 CP-parity,
$\vec{B}$ has to change sign when going from one jet to another. The sign of
the handedness correlation is completely determined at the parton level,
 but its
transfer to the hadron level is subject to the additional assumptions. We have
assumed that in the central rapidity region hadronization is blind to the
identity of the original parton (source of the jet), otherwise some
long range charge correlations would have been observed. In the fragmentation
region of the original quark (anti-quark), the process of hadronization of the
$(q\bar{q})_{s}$ pairs is strongly influenced by this identity. In an  extreme
case we could consider these processes as charge conjugate, which would reverse
the sign of the handedness correlation once again. For example, in the process
$e^{+}e^{-}\rightarrow R\bar{R}$, where R is a hadron resonance, the final
hadronic states of $R$ and $\bar{R}$ would be described by the charge
conjugate wave functions, hence, the charge conjugate of
 one and two particle dictributions. This would change sign in eq.(15).
 In fact such assumption is made
in papers [12], where authors claim the evidence for  the handedness
 correlations with the sign opposite to that expected by the standard QCD
 arguments.
\section{Chromomagnetic vacuum fields in hadron-hadron and nucleus-nucleus
collisions}
\par
Up to now we have discussed the effect of the chromomagnetic vacuum field on
the fragmentation of a single color string, such as we encounter in
$e^{+}e^{-}$ annihilation into hadrons. In hadron-hadron and nucleus-nucleus
collisions the domain of the chromomagnetic vacuum field can overlap with
many strings of different colors which fact complicates the situation,
but at the same time, may lead to stronger effects and richer and more
interesting phenomenology.
\par
As far as $\langle\Delta P_{\perp}\rangle$ and  $\langle P_{\perp}\rangle$
are concerned, multi-string situation amounts  to increase of $\overline{n}$
\--- the average number of $(q\bar{q})_{s}$ pairs per unit of rapidity. Thus
 a possibility arises to predict and study dependences of these quantities on
the number of color strings generated in the process. This number is
phenomenologically related
to the centrality of nucleus-nucleus collisions, to the number
of ``wounded quarks'',
inelasticities etc. The distributions of
$\Delta P_{\perp}$ and  $ P_{\perp}$ would be
broadened, and huge values of cumulant transverse momenta may arise due to
the  simultaneous fluctuations in $n$ and the random walk
distance.
\par
The effect on $R_{\Omega}$ of many strings with different colors  is more
complicated. Here the non-Abelian structure of the QCD theory comes into
play so that we have to add to the formula (15) an additional factor accounting
for a $SU(3)$-addition of the random color charges of strings. An accidental
color coherence between strings may lead to large  correlations of handedness.
In any case, this problem deserves a detailed analysis.
\par
The experimental requirement to observe the above mentioned effects is to have
a sign-of-charge-sensitive  measurement of the charged
 particles in the central rapidity region by a  precision tracker.
 It would be interesting to
investigate what are the possibilities of detectors constructed for RHIC
accelerator in this respect.
\section{Summary and conclusions}
We have seen, that simultaneous measurement of $\langle P_{\perp}\rangle$,
$\langle\Delta P_{\perp}\rangle$ and $R_{\Omega}$ permit,
within our  simple model,  an estimate of the vacuum
chromomagnetic field strength. It is certainly true, that our
purely classical treatment of
the interaction of quarks with the vacuum chromomagnetic field has not much
of a theoretical background.  Although some hints of the validity of such
phenomenology exist [13], a completely different picture of this interaction
leading to  very different results can be also considered [14]. We wish
to stress here our attitude of experimental physicists: let us
take the data and see what happens.
\par
The measurement we propose can be performed in the following, well defined
steps:
\begin{itemize}
\item
Select two-jet sample of the $e^{+}e^{-}$ events.
\item
Measure $\langle\Delta P_{\perp}\rangle$ as a function of
$\mid(y_{max}-y_{min})\mid$ and establish linear dependence (12) on
$\sqrt{\mid(y_{max}-y_{min})\mid}$ characteristic for the random walk in the
transverse momentum space.It should be noted that
the relation (12) reflects a property of the string fragmentation,
 independent
of the presence of the vacuum background chromomagnetic field.
 Its validity, with
some nonzero value of the parameter $\beta$ is a necessary condition for the
measurement we propose.
To obtain a non-zero result for  $\langle\Delta P_{\perp}\rangle$
we may have to  apply  an appropriate
cut on the minimal transverse momentum of particles, in order to diminish
the influence of the soft gluon emission, which is the main factor in diluting
 the charge correlations between partons and hadrons.
\item
Measure $\langle P_{\perp}\rangle$ as a function of $\mid y_{max}-y_{min}\mid$
and establish relation (7), which indicates the  presence of the transverse
momentum of the $(q\bar{q})_{s}$ pairs.
\item
Measure $R_{\Omega}$  and, using previously measured
$\langle\Delta P_{\perp}\rangle$, estimate the average value of the vacuum
 chromomagnetic  field strength.
\end{itemize}
\par
It is important to note, that measurements of $\langle P_{\perp}\rangle$
and $R_{\Omega}$ are independent, so we may choose $P_{\perp}$ as a "trigger"
for the events with large vacuum background field fluctuations. For example
we can select for further analysis only those two-jet events for which
$P_{\perp}$ exceeds by one or two standard deviations the average
$\langle P_{\perp}\rangle$
for both jets.
\par
Another important point is a  judicious choice  of the interval
 $(y_{min},y_{max})$ in the central
rapidity region for which the correlation $R_{\Omega}$ is determined.
It has to be as
large as possible, as the effect to be observed is proportional to the
interval length,
 but not too large in order to avoid trivial correlations due to the momentum
conservation.
\par
In an experiment with particle identification we could employ strong flavor
correlations between partons and hadrons. The $s\bar{s}$ pair should be
strongly correlated with $K^{+}K^{-}$ pair in the hadronic final state, so
constructions of $P_{\perp},\Delta P_{\perp} $ and $\Omega$ from $K^{\pm}$
momenta may be beneficial in spite of very much diminished statistics provided
that the effect of secondary interactions (different for $K^{+}$ and $K^{-}$)
can be taken into account.
\par
The absence of handedness correlation in SLD data [11] should not discourage
experimental physicists from the measurement we propose as it contains two
distinctive features:
\begin{itemize}
\item
 usage of the cumulative variables and possible enhancement of the effect
through random walk mechanism
\item
 possibility of "triggering" (through $ P_{\perp}$)
of large vacuum field fluctuations
\end{itemize}
\par
Most of the above remarks apply also to hadron-hadron and nucleus-nucleus
reactions, except for the complications introduced to multi-string situations
by the non-Abelian structure of theory. There we may expect a
rich and interesting phenomenology which may be of interest for the high energy
heavy ion experiments.
\par
It is clear that the measurements we propose are not easy , they demand
huge statistics and extremely careful checks of the systematic errors. On the
other hand the stake of this game \--- the experimental evidence for the
chromomagnetic component of the  vacuum \--- is worth to give  a try.
\vskip 1cm
\begin{center}
{\Large References}
\end{center}
\vskip 1cm
[1]G.K.Savvidy, Phys.Lett.{\bf 71B}(1977)133;\\ N.K.Nielsen, P.Olesen,
 Nucl.Phys.{\bf B144}(1978)376; \\ J.Ambjorn, P.Olesen,
Nucl.Phys.{\bf B170}(1980)265
\newline
[2]H.M.Fried, B.M\"{u}ller (eds.),"QCD Vacuum Structure", Proc. of the
workshop on QCD vacuum structure and its applications, Paris 1992
(World Scientific 1993)
\newline
[3]O.Nachtmann,A.Reiter, Z.Phys.{\bf C24}(1984)283
\newline
[4]G.W.Botz, P.Haberl, O.Nachtmann, Heidelberg University report
{\bf HD-THEP-94-36}; O.Nachtmann,{\bf HD-THEP-94-42}
\newline
[5]A.Di Giacomo, H.Panagopoulos, Phys.Lett.B285(1992)133;
H.D.Trottier, R.M.Woloshin, Phys.Rev.Lett.(1993)2053
\newline
[6]M.G.Ryskin,Phys.Lett.{\bf B319}(1993)346
\newline
[7]A.I.Vainstein,V.I.Zakharov,M.A.Shifman JETP Lett.{\bf 27}(1978)55
\newline
[8]H.G.Dosch,Prog. in Part. and Nucl.Phys. {\bf 33}(1994)121
\newline
[9]O.Nachtmann,Nucl.Phys {\bf B127}(1977)314
\newline
[10]A.Efremov, L.Mankiewicz, N.A.T\"{o}rnkvist, Phys.Lett {\bf B284}(1992)394;
 {\bf 291}(1992)473
\newline
[11]SLD Collaboration,K.Abe {\it et al.} Phys.Rev.Lett. {\bf 74}(1995)1512
\newline
[12]A.V. Efremov ,I.K.Potashnikova, L.G.Tkatchev, XXIX Rencontre de Morionds,
1994; A.V. Efremov ,I.K.Potashnikova, L.G.Tkatchev, 27 Int.Conf. on HEP,
Glasgow 1994,p.875;
A.V.Efremov,Puzzling correlation of Handedness in $Z^{0}\to 2 jet$ Decay;
I.K.Potashnikova, L.G.Tkatchev,Search for Jet Handedness in Hadronic
$Z^{0}$ Decays in ``Workshop on $\gamma\gamma$ Collisions'', Lund 1994,
World Scientific 1994;A.Efremov and D.Kharzeev, CERN-TH/95-139
\newline
[13]M.G.Ryskin,Sov.J.Nucl.Phys {\bf 48}(1988)708
\newline
[14]W.Czy\.{z},J.Turnau,Acta Phys.Polon. {\bf 24B} (1993)1501

\end{document}